\documentclass[aps,prd,preprint,amsmath]{revtex4}

\usepackage{graphicx}

\begin{document}

\title{ Big bang nucleosynthesis constrains the 
total annihilation cross section of neutralino dark matter }

\author{Xiao-Jun Bi}
\email{bixj@mail.ihep.ac.cn}
\affiliation{Key laboratory of particle astrophysics, IHEP, 
Chinese Academy of Sciences, Beijing 100049, P. R. China}
\affiliation{Center for High Energy Physics, Peking University,  
Beijing 100871, P. R. China}

\begin{abstract}

Assuming the lightest neutralino forms dark matter, we study
its residual annihilation after freeze-out at the early universe.
If taking place after the big bang nucleosynthesis (BBN) 
the annihilation products, especially at the hadronic modes, may
cause nonthermal nuclear reaction and change the prediction
of the primordial abundance of light elements in the standard BBN
scenario. We therefore put constraints on the neutralino annihilation
cross section. These constraints are free of the uncertainties
of the dark matter profile today suffered by direct or indirect 
detection of dark matter. We find the constraints by BBN is important,
especially when taking large $\tan\beta$. If the light element abundances
can be determined with higher precision in the future the constraint will 
become very strong, so that a majority of the parameter space allowed by the 
relic density requirement may be excluded.

\end{abstract}

\maketitle

The existence of cosmological dark matter (DM) has been firmly established by
a multitude of astronomical observations.
However, the nature of the non-baryonic dark matter is still unknown
and remains one of the most outstanding
puzzles in particle physics and cosmology.
Among a large amount of theoretical candidates, the most attractive
scenario involves the weakly interacting massive particles (WIMPs).
An appealing idea is that the WIMPs freeze out at the very early time and 
form the thermal relics,
which naturally account for the relic abundance 
observed today \cite{wmap}. The WIMPs are well motived
theoretically in particle physics beyond the standard model to solve the
hierarchical problem. 
In particular, the minimal supersymmetric extension of the standard model
(MSSM) provides an excellent WIMP candidate as the lightest supersymmetric
particle, usually the lightest neutralino, which are stable due to R-parity
conservation \cite{jungman}. 
%The cosmological constraints on the
%supersymmetric (SUSY) parameter space have been extensively studied in
%the literature \cite{csusy}.

The WIMPS can be detected
on the present running or future experiments, either directly
by measuring the recoil energy when WIMP scatters off the
detector nuclei \cite{direct} or indirectly by observing the
annihilation products of the WIMPs,
such as the antiprotons, positrons, $\gamma$-rays or
neutrinos \cite{jungman,indirect}.
After decades of efforts, the sensitivity of these experiments have
been improved by many orders of magnitude.
However, no positive signals have been found up to now.

Conversely, the null results put constraints on the parameter
space of the dark matter model, such as the MSSM.
However, all the WIMP detection experiments depend on the
distribution profile of dark matter.
Especially for the indirect detection the predicted annihilation
products from the Galactic Center (GC) \cite{gc}
can vary for several orders of magnitude by assuming
different dark matter profiles.
In theoretical studies, in order to give optimistic predictions
a cuspy dark matter profile is usually adopted, 
such as the NFW \cite{nfw} or Moore
\cite{moore} profile which is favored by N-body simulation.
However, observations of rotation curve strongly disfavor cuspy profiles.
Instead, they generally favor a cored profile \cite{obser}.
The discrepancy between simulation and observation has been thought 
a severe challenge to the cold dark matter scenario for cosmological structure
formation.
If a cored profile is adopted the theoretical prediction of dark matter
annihilation (DMA) products from the GC may be below the sensitivities of
all present or near future experiments. 
Therefore, no firm constraints can be set on the 
MSSM parameter space from the present dark matter detection
experiments. 

More firm constraints on the dark matter model may come from
the early universe processes when the density fluctuation is very small.
Indeed, the most stringent constraint on the MSSM parameter space today
actually comes from the process of the lightest supersymmetric particle
(LSP) decoupling at the early universe,
by requiring the relic density of the LSP being consistent with
the measurement of WMAP \cite{csusy}.
Besides that
there are also model independent constraints on the dark matter 
annihilation cross section from unitarity bound \cite{gk,hui}
or from measurement of the cosmological neutrino flux \cite{beacom},
which, however, set much loser constraints than that given by 
the decoupling process.

In the present work we will set a new constraint on the MSSM
parameter space from another process at the early universe,
i.e. the big-bang nucleosynthesis (BBN).
Knowing the nuclear reaction processes and the 
evolution history of the universe in the standard cosmology 
we can precisely calculate the abundances of the light elements,
mainly on D, $^3$He, $^4$He, $^6$Li, $^7$Li.
The standard BBN scenario gives
consistent predictions of light element abundances compared with observations.
The agreement between the BBN predictions and observations
can be used to constrain various processes beyond the
standard cosmology or standard particle physics.
For example, the BBN has been extensively studied in the literature 
to constrain the long-lived
heavy particles, such as gravitino \cite{decay}, which may decay
after BBN.

We will  investigate how the standard
BBN constrains the neutralino self-annihilation.
%annihilation may have an impact on the successful
%predictions of the standard big-bang nucleosynthsis.
After neutralino freeze-out there continues to be some residual 
annihilation of neutralinos.  
Although rare on the expansion time scale the residual self-annihilation
continues to produce high energy particles well after BBN ends,
changes the abundances of light elements,
thereby ruin the agreement between BBN theory and observations.
Therefore observational data of light element abundances constrains the rate of
neutralino self-annihilation.
It should be noted that since the rate of the WIMP annihilation 
is proportional to the number
density square of the dark matter particles, at the early universe
the annihilation rate of neutralino is much higher than the
average rate today.
% despite the rate is too small compared with the
%unverse expansion rate after
%freeze-out and have no impact on the thermal bath. 
%Our calculation shows that the dark matter annihilation (DMA) after BBN 
%may leave some imprints on BBN and therefore constrain the annihilaton
%rate of DMA. 
%For this reason BBN can be used to constrain the interaction strength
%and mass of WIMPs. 

The abundances of light elements are especially sensitive to the
injecting of strongly interacting particles during 
nucleosynthesis.
The main effect of the hadronic cascades is
that the ambient $^4$He is destroyed and D, T, $^3$He and  $^6$Li, $^7$Li
are created.
%Since the nonthermally produced D, $^3$He, $^6$Li and $^7$Li change the
%prediction of standard BBN one might expect severe constraint on the 
%hadronic injection from the neutralino self-annihilation.
BBN with hadronic-dissociation processes induced by hadronic decays 
of long lived X-particles (any theoretical assumed long lifetime particles) 
was studied in 
\cite{decay}. %especially for the case of gravitino decay.
In this work we study the effect of injecting hadronic particles from
neutralino annihilation on BBN.

%Limits on the self-annihilation cross section and mass of Xparticle have been
%given by requireing that electromagnetic cascades generated by the
%self-annihilation of X do not jeopardize the prediciton of the bigband
%nucleosysnthesis.

To derive the constraint we will follow the calculation given by 
M. Kawasaki et al. \cite{decay} closely. In \cite{decay} the authors 
adopted the most recent data of nuclear reaction cross sections and
observational light element abundances, new Monte Carlo
event generator for quark/gluon hadronization. The evolution of
the hadronic shower in the thermal bath is also carefully treated.
Taking the uncertainties of the measurements into account 
quite conservative constraints on the abundances of X-particles as
a function of its life time are derived. 

The X-particle is assumed to have a two-body decay into monoenergetic quarks
with energy $E=m_X/2$,
which evolve into two jets. The injection rate of the jets is determined
as $n_X/\tau_X$, with $n_X$ and $\tau_X$ the number density and
life time of X-particle respectively. Finally the constraint
on the relative number density of $X$, $Y_X=n_X/s$ with $s$ the entropy
of the Universe, as function of its lifetime $\tau_X$ is given.
Considering that neutralino annihilation
also produces  monoenergetic quarks we can roughly relate the
constraints on the injection rate $n_X/\tau_X$ in \cite{decay}
into the neutralino annihilation
rate, which % The injection rate of jets from neutralino annihilation
is determined by 
\begin{equation}
\label{rate}
R=\frac{<\sigma v>}{2} n_\chi^2\ ,
\end{equation} with $<\sigma v>$ the thermal
averaged annihilation cross section and 
$n_\chi$ the number density of neutralino.
The factor $2$ is due to identical particles of initial state.

The density of dark matter is given as $\rho_{DM} = \rho_{DM}^0 a^{-3}$ with
$\rho_{DM}^0 $ the dark matter density today and 
$a$ the cosmological scale factor.
The entropy $s$ of the Universe is given in the same way.
The time is related with the scale factor by
\begin{equation}
\label{time}
t = \int_0^a \frac{d a}{\dot{a}} = \int_0^a \frac{d a}{ a H} = 
\frac{1}{H_0}  \int_0^a \frac{d a}{a \sqrt{\frac{\rho}{\rho_0}}}\ ,
\end{equation}
where the Hubble constant is related with that today by 
$\frac{H^2}{H_0^2} = \frac{\rho}{\rho_0}$.
The energy density is given by $\rho = \rho_\gamma^0 a^{-4}+
\rho_\nu^0 a^{-4}+ \rho_m^0 a^{-3} +
\rho_\Lambda^0$, with $\rho_\gamma^0$, $\rho_\nu^0$, $ \rho_m^0$
and $\rho_\Lambda^0$ the radiation, neutrino, matter and dark energy
density today.

\begin{figure}
\includegraphics[scale=0.9]{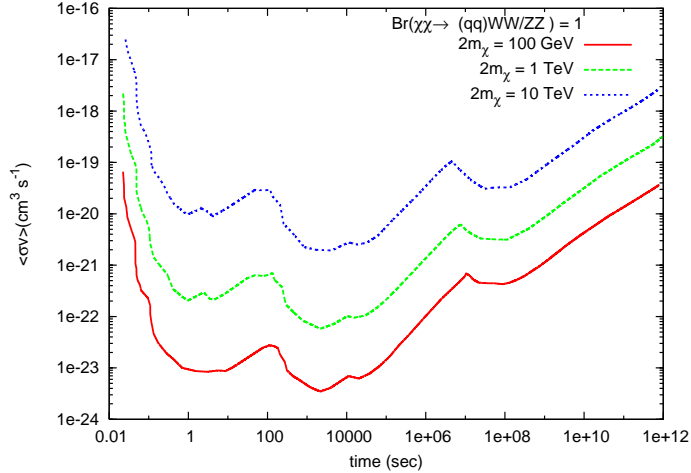}
\caption{\label{sv}
Constraints on $<\sigma v>$ as function of time $t$ for  
$2m_\chi = 100 GeV, 1 TeV,\text{and}\ 10 TeV$. we have 
assumed that the neutralino
annihilates totally into gauge bosons $W^+W^-/ZZ$ for $m_\chi > m_W$
or quarks for $m_\chi < m_W$.
}
\end{figure}

Taking the initial conditions of $\rho^0_i$ and integrating Eq. (\ref{time})
we get the number density and therefore the annihilation rate 
$R$ of neutralino at any time $t$. 
%The reaction of the hadronic jets injected at different time is also different
%and the constraints on the injection rate varies with time.
From the constraints on the hadronic jets injection rate derived
in \cite{decay}  and Eq. (\ref{rate}) we get constraints 
%on the hadronic jet injection rate from
%neutralino annihilation as function of time $t$ from the constraints
%on $X$ particle decay in \cite{decay}, which leads to the constraints 
on the annihilation cross section $<\sigma v>$ at $t$. 
An accurate result should be given by solving the Boltzmann equation
numerically.
Here we give constraint by a simply correspondance betwen
X-particle decay and neutralino annihilation. This may lead to a
conservative constraint on the annihilation cross section.
%due to that at early time than $t$ the annihilation
%rate is higher than X decay rate.
In Fig. \ref{sv} we show the constraints on $<\sigma v>$ as function
of time $t$ for $2m_\chi = 100 GeV, 1 TeV,\text{and}\ 10 TeV$ respectively.
The constraints are corresponding to these constraints 
for $X$ decay with $m_X=100GeV, 1 TeV,
\text{and}\ 10 TeV$ respectively \cite{decay}. 
In deriving the constraints, we have assumed that the neutralino 
annihilates totally into gauge bosons $W^+W^-/ZZ$ for $m_\chi > m_W$, 
where the gauge
bosons decay to quarks with the branching ratio $70\%$.
We have checked that this assumption is quite reasonable.
%for most SUSY models. 
For neutralinos 
above the threshold $m_\chi > M_W$ they annihilate dominantly into gauge bosons
in most MSSM models.
For $m_\chi < m_W$ we assume neutralino annihilates totally into quarks.
(In the following Fig. \ref{model} we will give constraints on the MSSM
parameters by directly calculating the annihilation products without
any assumptions.)

\begin{figure}
\includegraphics[scale=0.9]{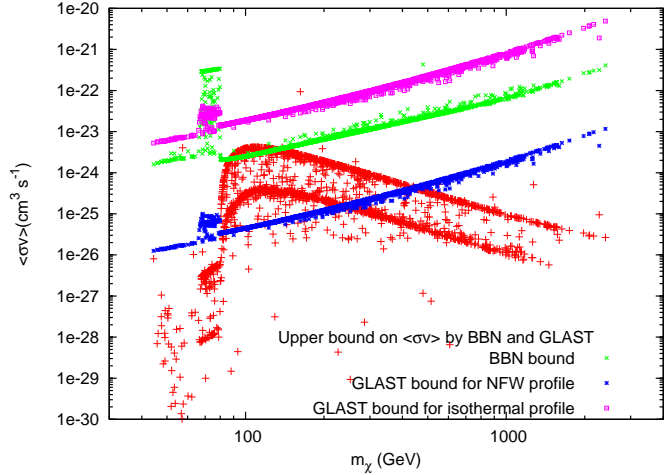}
\caption{\label{model}
The constraints on $<\sigma v>$ in the MSSM parameter space set from
BBN and GLAST by observation of DM annihilation at the GC.
For the constraints by GLAST two DM profiles, NFW and isothermal,
are adopted. 
}
\end{figure}

From Fig. \ref{sv} the strongest constraints on $<\sigma v>$ is
at $t\approx 2000$ sec, which come from the $^6$Li/H data.
The abundance of $^6$Li is very sensitive to
the nonthermal hadronic jet injection. However, the $^6$Li abundance
is difficult to determine. The standard BBN prediction of $^6$Li
abundance is $(^6Li/H)_\text{SBBN}=1.30\times 10^{-14}$. Taking the
large uncertainties in determining  $^6$Li abundance the constraints
on hadronic jet injection is given by assuming  
$(^6Li/H) < 10^{-11} \sim 10^{-10}$, which is several orders of magnitude
higher than the standard prediction.
Therefore the constraints on the nonstandard process from BBN
can be much stronger if $^6$Li can be determined with higher precision.

Adopting the constraints on $<\sigma v>$
at $t\approx 2000$ sec we show how the MSSM parameter space is constrained
by BBN in Fig. \ref{model}. 
The dots in the figure are produced randomly in the MSSM parameter
space and the corresponding $<\sigma v>$ is calculated using the
package DarkSUSY \cite{darksusy}. The constraints on $<\sigma v>$
for different neutralino mass is given by interpolation of the constraints
for $2m_\chi =100 GeV, 1 TeV, 10 TeV$. The `BBN bound' in the figure
shows the constraints on $<\sigma v>$ by BBN. The scatter of the
bound comes from the different branching ratios of neutralino annihilation
to quarks. Near the threshold neutralino may annihilate into leptons dominantly.
In such cases we take the constraints from $^3$He/D which is sensitive
to the photodissociation process. The
most stringent constraint from the photodissociation process is given
at $t \approx 8.5\times 10^7$ sec.

In Fig \ref{model} we also show the parameters
that can be detected by GLAST \cite{glast}. In theoretical prediction
of dark matter annihilation we usually adopt the dark matter profile
from N-body simulation, which generally predicts cuspy profiles such
as NFW \cite{nfw} or Moore \cite{moore} profiles. The NFW or Moore
profiles have singularities at the halo center
as $\rho_\text{NFW}\to r^{-1}$ and $\rho_\text{Moore}
\to r^{-1.5}$ respectively. The singularity leads to large (or divergent)
annihilation flux and can be detected by the satellite detectors, such
as GLAST. However, observation of rotation curves
usually strongly favor a cored dark
matter profile, instead of cuspy ones \cite{obser}. 
If adopting a cored dark matter
profile the present detectors will have much weaker potential to
detect the signals from dark matter annihilation. In Fig. \ref{model}
we show the constraints on $<\sigma v>$ from observation of
DM annihilation at the GC by GLAST, assuming both NFW and cored profiles.
In deriving the constraints by GLAST we take the gamma ray source detected
by HESS at the GC \cite{hess} as background and extend it to lower energy.
The present bound by BBN has been stronger than that 
set by GLAST taking a cored profile, while weaker if taking 
a NFW profile.

\begin{figure}
\includegraphics[scale=0.8]{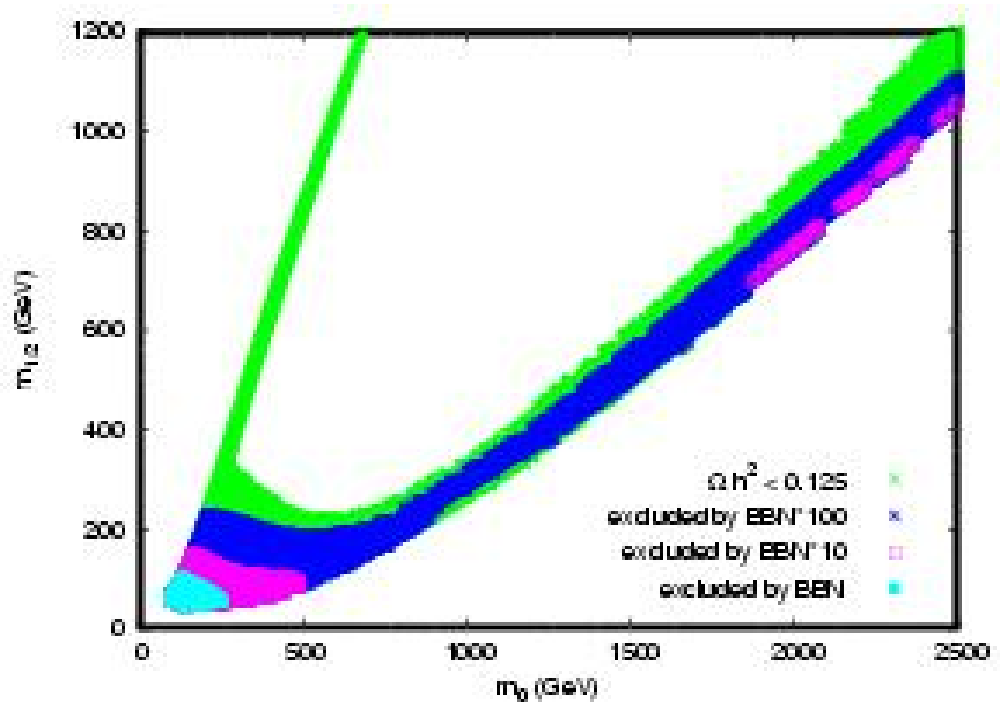}
\includegraphics[scale=0.8]{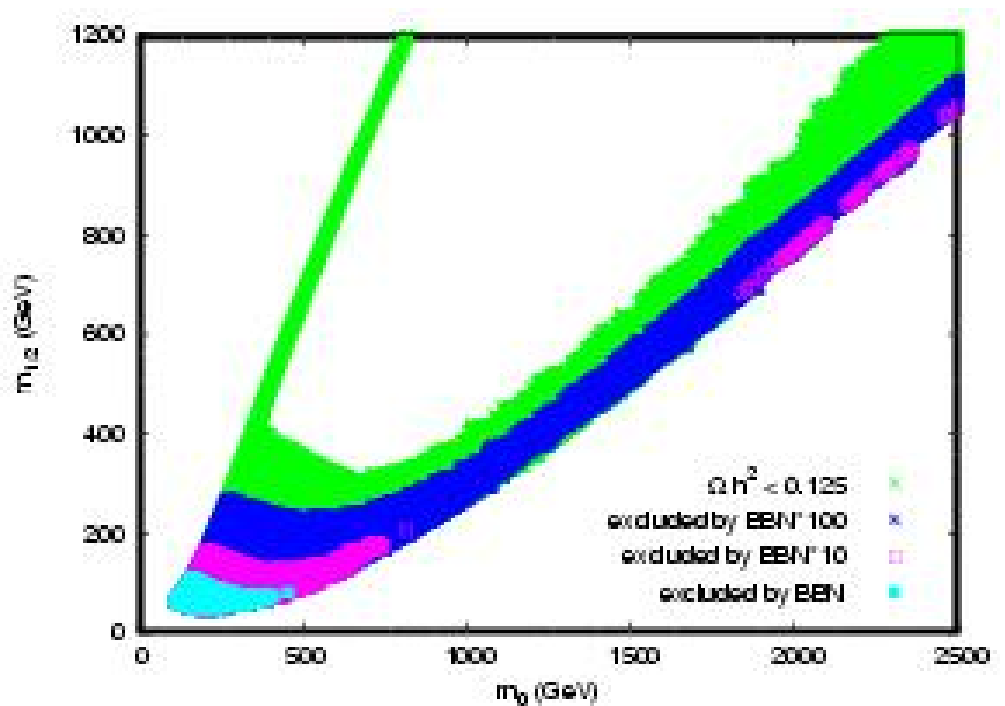}
\includegraphics[scale=0.8]{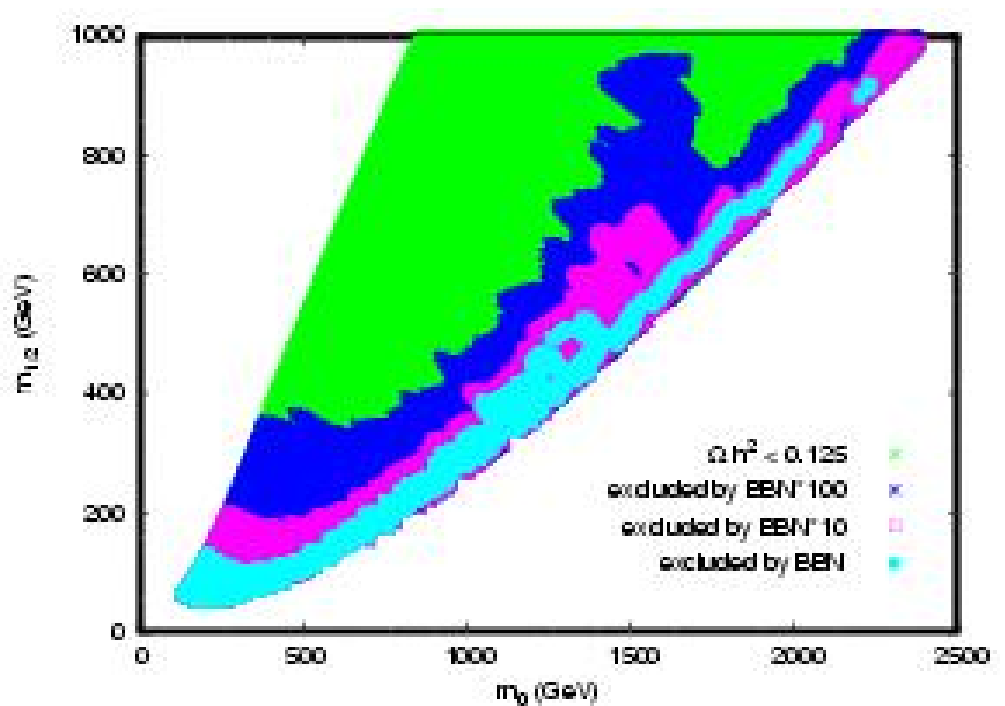}
\caption{ \label{msugra}
The parameter space that satisfies $\Omega_ch^2 < 0.125$
and these excluded by BBN constraints. `BBN*10' and `BBN*100'
means how the parameter space is constrained if the precision of $^6$Li data
is improved by one and two orders of magnitude.
}
\end{figure}

Further, we set the exclusion region by BBN in the parameter space
of the minimal super-gravity mediated SUSY breaking model (mSUGRA). 
In calculating the relic density of mSUGRA models, the
package MicrOMEGAs 2.0.7 is adopted \cite{micro}, where 
the package ISAJET \cite{isajet} is incorporated to run the 
renormalization group equations 
from the GUT scale to the low energy scale.
The BBN bounds are especially important when taking large values of
$\tan\beta$.
In Fig. \ref{msugra} we show the exclusion region on the $m_0-m_{1/2}$ plane
set by BBN taking
$\tan\beta=50, 55, \text{and}\ 60$ respectively. We have taken $A_0=0$,
and $\mu$ positive. In the  $m_0-m_{1/2}$ plane the whole
shaded region represents
the models which satisfy the WMAP \cite{wmap} constraints on the
relic density. The WMAP 5-year data gives 
$\Omega_c h^2 = 0.1143\pm 0.0034$ \cite{wmap}.
The shaded region in Fig. \ref{msugra} is given by requiring the CDM relic
density be smaller than the $3\sigma$ upper bound, 
i.e., $\Omega_c h^2 < 0.125$.
The excluded region marked as `BBN*10' and `BBN*100' represent that the
precision of $^6$Li data is improved by 10 and 100 times in the future
respectively.
For $\tan\beta = 60$ the present BBN bound has excluded a large part
of the allowed models. For  $\tan\beta = 50,\ 55$ only these models
with small values of $m_0$ and $m_{1/2}$ are excluded by BBN.
If the $^6$Li data is improved by 2 orders of magnitude we can see
that most of the parameters allowed by WMAP will be excluded.

It should be noted that we only take the upper bound of the relic
density from WMAP into account. That means the neutralino may only
account for a part of dark matter density, or there are nonthermal
contribution to the relic density \cite{nonthermal}. Therefore
the present BBN bound set constraints only for the large 
annihilation cross section. However, as have seen, it is even
more severe than that set by GLAST for a cored profile. 
Although the neutralino decoupling process gives the most
stringent constraint now, it can be changed in nonstandard cosmology
as shown by Gelmini and Gondolo \cite{nons} since the process takes
place at very early time when we know very little.
However, the bound from BBN is much solid and hard to invalidate it.
BBN and cosmic microwave background have long been taken
as two classic proof of the success of the standard cosmology.
Compared with other model independent bound on the DM annihilation rate 
\cite{gk,hui,beacom} the present bound is much more severe.
With the improvement of the precision of light element abundances 
the exclusion bound can also be greatly improved. 
Anyway, we present a new constraint on the MSSM parameter space  
independent of the N-body simulation result,
besides that from the decoupling process.

In summary, in this work we study how the residual annihilation of
neutralino after freeze-out can affect the abundance of light elements 
predicted in the standard scenario. According to the study we try to
set constraints on the SUSY parameter space. The constraints are
different from these set by direct or indirect detection of dark matter
which heavily depends on the dark matter profile. The dark matter profiles
are usually predicted by N-body simulations, which, however, seem to show
discrepancy with the observation of rotation curves. 
This has been taken as a serious
problem in structure formation in the cold dark matter scenario.
Our result shows that BBN can give quite strong constraints on the
SUSY parameter space.
% compared with other model-independent constraints\cite{gk,hui,beacom}.
Especially, the most stringent constraint comes from the $^6$Li data which,
however, has very large uncertainties. The present constraint is given by
requiring that the $^6$Li abundance is lower than $10^3 \sim 10^4$ times 
the prediction of the standard scenario. If the bound can be improved by
2 orders of magnitude we find a large part of the 
important SUSY parameter space will be excluded by BBN.

\begin{acknowledgments}
This work is supported by the NSF of China under the grant
Nos. 10575111, 10773011 and supported in part by the Chinese Academy of
Sciences under the grant No. KJCX3-SYW-N2.
\end{acknowledgments}

\end{document}